# A Measurement Method of Diffusion Coefficient of Liquid Using Radial Laser Rays Formed By Cylindrical Refractive System

Ju Dok-Yong, Jo Jong-Hyon, Kim Nam-Chol*

Faculty of Physics, **Kim Il Sung** University

Pyongyang, Democratic People's Republic of Korea

*Email Address: ryongnam19@yahoo.com

**Abstract**

The precision measurement of diffusion coefficient of solution is very important for the clear understanding of material transfer and interaction between the materials. But the high precision measurement of diffusion coefficient is very difficult compared to that of the other physical quantities. In this paper, we have proposed a new method to determine the diffusion coefficient by using the radial laser rays formed by cylindrical refractive system (cylindrical glass rod) and **CCD** camera. The cylindrical glass rod is placed in front of laser device and then, the laser rays spread radially. The slant curve with normal distribution profile is observed by illuminating these rays to the diffusion layer boundary. This curve is recorded on **PC** by **CCD** (charge coupled device) camera and the shape of this curve varies with time. The variation of maximum value with time in the curve determines diffusion coefficient. The measured diffusivity value is in good agreement with the previous experimental result in the tendency.

## 1. Introduction

Diffusion is a physical phenomenon which describes the transfer of material particles to the zones of lower concentration from the zones of higher concentration by the concentration gradient. This phenomenon first was discovered in the nineteenth century. Since then, it has attracted great interests not only because it shows the molecular behavior at the atomic level but also due to its increasing applications in the fields such as chemical industry, foodstuff industry, physiology and medicine [1]. It is a basic problem to grasp the interaction between the materials and the transport phenomenon of particles [2, 3]. Diffusion coefficient is an important index for the quantitative understanding on the diffusion process. Therefore, researches for measuring its value are widely carried out. There are many techniques to find diffusivity; for example, total internal reflection fluorescence (**TIRF)** microscopy [4]**,** spectroscopy [5], optical interferometric techniques [6, 7] and pulsed field gradient **NMR** [8].

**TIRF** microscopy uses **TIRF** microscope to record the position of fluorescence as a function of time and thus can determine the diffusion coefficient of fluorescent molecules.

This method is perhaps simple but has the fault limited to fluorescent substance. The other methods measure the concentration change as a function of time and reveal high accuracy. Especially, holographic interferometric technique is the most potential one [9-14]. But as indicated in the literature [15], the main fault of this technique is that the optical requirement is too strong and as a result, it is difficult to realize the industrial application. It is because spectroscopy or interferometry including holographic interferom etry requires complicated experimental setup and procedures

Recently, a method using Michelson Interferometer is proposed to measure the diffusion coefficient of transparent solution [15-17]. Though this technique does not need recording medium and the shift in the interferogram is easy identifiable, there are some drawbacks, i.e., the technical requirement for the optical system to measure the diffusivity is stringent and the measurement method is complicated.

In the present paper we proposed a much simpler method for determining the diffusion coefficient of solution by using radial laser rays formed by cylindrical glass rod and **CCD** camera.

## 2. Experimental set up

The experimental configuration for the measurement of the diffusion coefficient is shown in Fig.1.

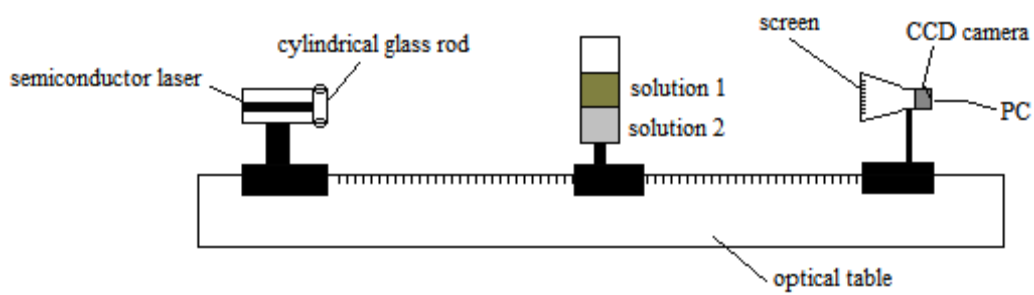

Fig.1. Configuration of experimental setup.

The semiconductor laser used in the experiment is red one (output 5mW, wavelength 635nm). The cylindrical glass rod with the diameter of 0.1mm is placed in front of laser source. **PC** camera with the pixels of 480×620 is used as a **CCD** camera and it is connected with computer in USB communication manner. There is a screen to receive an interested image before **CCD** camera. The light from a laser device is spread radially passing through the cylindrical glass rod as shown in Fig. 2(a). As the radial laser rays are incident on the cuvette containing homogeneous solution, the line image can be obtained on the screen (Fig.

2(b)). However, in case of the incidence on the interface (diffusion boundary layer) of different two solutions, the slant normal distribution pattern can be occurred (Fig.2(c)). Here, Fig. 2(b) is for the distilled water with the refractive index of 1.333 and Fig.2(c) for the mixing of the distilled water and the salt water of concentration 30%, of which refractive index is 1.386. The distilled water is first put into the cuvette and then the salt water of density greater than the distilled water is allowed to sink below the distilled water using the injector.

In this picture maximum deviation H is gradually reduced with the time during the diffusion process and after the passage of a little time, the image is transformed into a line. This is due to the reduction of concentration difference of two solutions and hence the reduction of refractive index difference with the time during the diffusion around the contact of two solutions. Therefore, the diffusivity of solution can be determined by measuring accurately the variation property of this maximum deviation. Conclusively, in the paper we have proposed a new method for the determination of diffusion coefficient, which is to measure the variation of maximum deviation corresponding to the extreme point with the time in the slant curve.

## 3. Theory

### 3.1. Relation between the diffusion coefficient and the change of refractive index of solution

Generally, the one dimensional free diffusion process is given by Fick's second law [9, 16]:

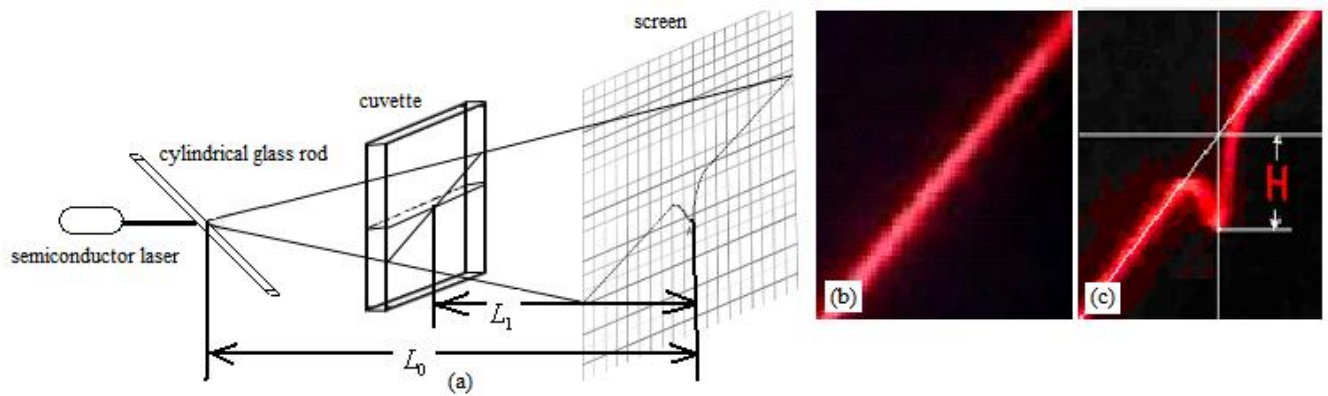

Fig.2. Operation principle of measurement device and image of laser beam on the screen: (a) Operation principle; (b) after passing through cuvette containing homogeneous solution; (c) after passing through the diffusion boundary layer.

$$\frac{\partial C(x,t)}{\partial t} = D \frac{\partial^2 C(x,t)}{\partial x^2}. \tag{1}$$

Where $C(x,t)$ is the concentration at any position $x$ at any time $t$ for one dimensional diffusion along $x$-axis, D is the diffusion coefficient of solution. The generalized solution of this equation is already known as

$$\frac{\partial C(x,t)}{\partial x} = \frac{n_2 - n_1}{2\sqrt{\pi Dt}} \exp\left(-\frac{x^2}{4Dt}\right). \tag{2}$$

This shows the change of concentration gradient around the diffusion boundary. Here $n_1$ and $n_2$ are the initial refractive indices of two solutions at the time t=0.

On the other hand, the relationship between the concentration and refractive index can be written as [9]

$$n(x,t) = mC(x,t) + n_0. \tag{3}$$

Here $m$ is the slope of curve between concentration and refractive index as a constant and $n_0$ is a constant. Therefore, substituting Eq. (3) into Eq. (1) and Eq. (2), the following equations can be obtained as

$$\frac{\partial n(x,t)}{\partial t} = \frac{\partial^2 n(x,t)}{\partial x^2}, \tag{4}$$

$$\frac{\partial n(x,t)}{\partial x} = \frac{n_2 - n_1}{2\sqrt{\pi Dt}} \exp\left(-\frac{x^2}{4Dt}\right). \tag{5}$$

These equations depict the change of refractive index gradient around the diffusion boundary. Eq. (5) can be rewritten as

$$\frac{\partial n(x,t)}{\partial x} = \frac{\Delta n}{\sqrt{2\pi}\sigma} \exp\left(-\frac{x^2}{2\sigma^2}\right). \tag{6}$$

Where $\sigma = \sqrt{2Dt}$ and $\Delta n = n_2 - n_1$. Eq. (6) is similar to the normal distribution function.

Fig.3 illustrates the change property of refractive index gradient around the diffusion boundary. Solution 1 and Solution 2 contact together each other at the origin 0 of $x$ axis. Just the origin is the diffusion boundary. The scale of $x$ axis is the relative scale calibrated at the diffusion boundary.

Eq. (5) has the maximum value at $x = 0$; i.e.,

$$\left(\frac{\partial n(x,t)}{\partial x}\right)_{max} = \frac{n_2 - n_1}{2\sqrt{\pi Dt}}. \tag{7}$$

The maximum value of refractive index gradient is shown in Fig. 3. From Eq. (5), the

diffusion coefficient can be defined as

$$D = \frac{(n_2 - n_1)^2}{4\pi t \left(\frac{\partial n(0,t)}{\partial x}\right)^2_{max}}. \tag{8}$$

Therefore, if the maximum value of refractive index gradient is measured, the diffusion coefficient can be determined.

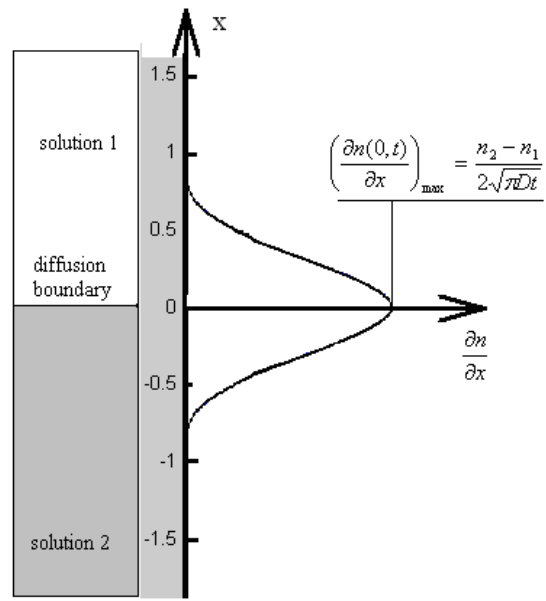

Fig.3. Change property of refractive index gradient in the solution

### 3.2. Relation between the diffusion coefficient and the deviation angle of refractive ray

The maximum value of refractive index gradient can be determined through the measurement of the maximum deviation angle of refractive ray passed through the solution layer. The refractive angles of rays are different with the position due to the difference of refractive index of solution with the position as in Fig.4 when the parallel rays are incident perpendicular to the diffusion boundary, at which two solutions with the initial refractive indices $n_1$ and $n_2$ contact.

Refractive angle of each ray can be calculated using the principle of elementary wave interference and refractive law.

$A_1$ and $A_2$ are any adjacent two locations on the wave front of plane wave illuminated perpendicularly to the cuvette, respectively. After the time interval $\Delta t$ with the beginning of diffusion, the wave front will be determined by the envelope $B_1 B_2$ surrounded the elementary waves as shown in Fig.5. Suppose that the refractive index of solution at the location $A_2$ is $n(x,t)$ and $n(x,t) + \Delta n(x,t)$ at the location $A_1$.

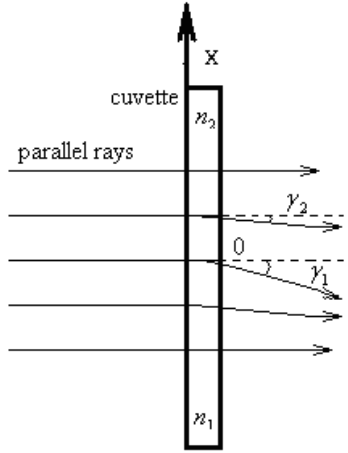

Fig.4. Refractive angle of parallel rays incident into solution

The propagation velocity of the light is $c/(n(x,t)+\Delta n(x,t))$ at the point $A_1$, and then, $c/n(x,t)$ at the point $A_2$. Here, $c$ is the velocity of light in vacuum and $n$ is the refractive index of the solution. Thus, the radius of elementary wave propagating at the point $A_2$ is

$$A_2 B_2 = \frac{c\Delta t}{n(x,t)}. \tag{9}$$

And the radius of elementary wave propagating at the point $A_1$ is

$$A_1 B_1 = \frac{c\Delta t}{n(x,t)+\Delta n(x,t)}. \tag{10}$$

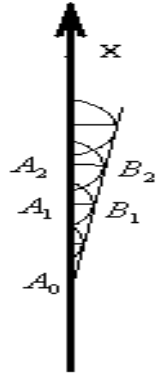

Fig.5. Schematic diagram of elementary wave interference

From Fig. 5, the following relation is obtained as

$$A_0 A_1 = A_1 A_2 \frac{A_1 B_1}{A_2 B_2 - A_1 B_1}. \tag{11}$$

with $A_1 A_2 = \Delta x$. Substituting Eq. (9) and (10) into Eq. (11) and rearranging Eq. (11), the simplified relation can be found as

$$A_0 A_1 = n(x,t)\frac{\Delta x}{\Delta n(x,t)}. \tag{12}$$

And at the limit of $A_2 \to A_1$ (or $\Delta x \to 0$), $A_0 A_1$ converges to the radius of curvature of the wave front at the point $A_1$, $R$, i.e., $\lim\limits_{\Delta x \to 0} A_0 A_1 \to R = \dfrac{n(x,t)}{\partial n(x,t)/\partial x}$.

The deviation of ray in the solution and air is shown in Fig.6. In the figure, $\delta$ is the thickness of solution layer, $\alpha$ is the angle deviated from the original direction of propagation, and $\beta$ is the deviation angle at the moment passing through the air from the solution. For enough thin thickness of solution layer, the deviation angle of light during the passing through the solution is given by

$$\alpha = \delta/R. \tag{13}$$

At the solution-air boundary, $\alpha$ is an incident angle and $\beta$ is a refractive angle. If both angles are very small in the magnitude, the refractive law can be written as

$$\beta(x,t) = n(x,t)\alpha = n(x,t)\frac{\delta}{R} = \delta\frac{\partial n(x,t)}{\partial x}. \tag{14}$$

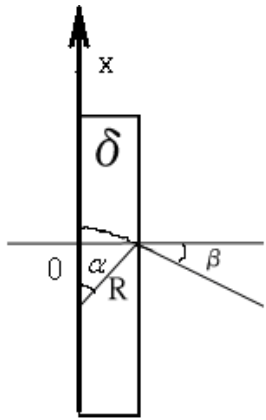

Fig.6. Deviation of ray in the solution and air.

Substituting Eq. (4) into Eq. (14), the deviation angle of ray passed through the solution layer around diffusion boundary can be expressed as

$$\beta(x,t) = \frac{\delta \cdot (n_2 - n_1)}{2\sqrt{\pi Dt}} \exp\left(-\frac{x^2}{4Dt}\right). \tag{15}$$

As can be seen above, the refractive angle is the maximum at the position ($x = 0$) that the variation rate of refractive index is the largest, and thus the maximum deviation angle can be expressed as

$$\gamma = \beta(0,t)_{max} = \frac{\delta \cdot (n_2 - n_1)}{2\sqrt{\pi Dt}}. \tag{16}$$

As a result, the maximum of refractive index gradient is given by

$$\left(\frac{\partial n}{\partial x}\right)_{max} = \frac{\gamma}{\delta}. \tag{17}$$

The distribution of deviation angle with the position is shown in Fig.7.

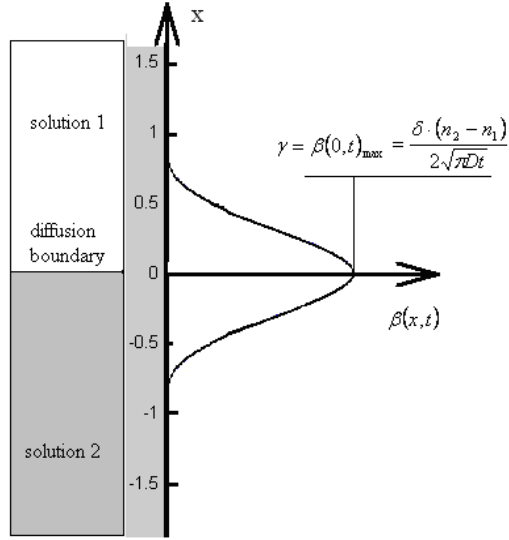

Fig.7. Distribution of deviation angle with the position

Substituting Eq. (17) into Eq. (8), the diffusion coefficient can be written as

$$D = \frac{\delta^2 (n_2 - n_1)^2}{4\pi t \gamma^2}. \tag{18}$$

Thus, to find the diffusion coefficient, it is sufficient to measure only the maximum deviation angle $\gamma$ at the time $t$ for the known $n_1$ and $n_2$. However, in the practice, $n_1$ and $n_2$ should be inevitably measured in many cases. If the maximum deviation in the image on the screen, $f_{max}(x) = H$ in Fig.3-(c) is very small compared to the distance $L_1$ between cuvette and screen ($L_1 \gg H$), maximum deviation angle is given by

$$\gamma = \frac{H}{L_1}. \tag{19}$$

Thus, the diffusion coefficient is expressed as

$$D = \frac{\delta^2 (n_2 - n_1)^2}{4\pi t \gamma^2} = \frac{\delta^2 (n_2 - n_1)^2 L_1^2}{4\pi t H^2}. \tag{20}$$

Generally, the extreme points in Fig.4 and Fig.8 are not accurately coincided with the position of extreme point in Fig.3(c), and hence, the minimum value (H) in Fig.3(c) itself cannot be utilized.

## 3.3. A method for the measurement of maximum deviation angle

When radial laser rays are incident perpendicularly to diffusion boundary (parallel to OX-axis), the incident and refractive rays are superposed. It reveals a line on OX-axis, and thus, it is difficult to distinguish them. As a result, the maximum deviation angle of refractive ray cannot be measured.

On the other hand, when radial laser rays are incident parallel to diffusion boundary (OY-axis), a broad line compared to incident ray is formed on the screen by refractive index change around diffusion boundary. However, the variation of maximum deviation angle cannot be accurately measured as its edges are obscure. If the radial ray is incident to the diffusion boundary with a constant slant angle, the following image can be obtained (see Fig.8).

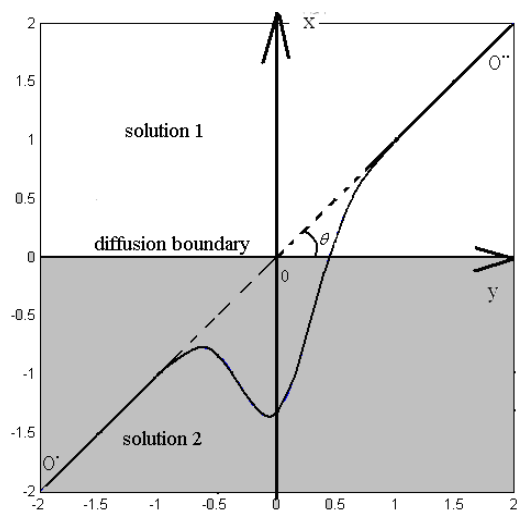

Fig.8. Image obtained by ray incident slant to diffusion boundary

As it can be seen in Fig.8, the maximum deviation angle at x=0 is always constant even though the slant angle of $O'O''$ axis changes. The distribution of deviation angle with respect to $O'O''$ axis depends on the Eq. (15), which describes the distribution of deviation angle according to x-axis, and slant ray along $O'O''$ axis. Equation of $O'O''$ axis can be expressed as

$$f(x) = Ax . \qquad (21)$$

Where $A = \tan\theta$. Therefore, the distribution of deviation angle on the $O'O''$ axis can be obtained from Eq. (15) and Eq. (21), i.e.,

$$\beta'(x,t) = \frac{\delta \cdot (n_2 - n_1)}{2\sqrt{\pi Dt}} \exp\left(-\frac{x^2}{4Dt}\right) + Ax. \qquad (22)$$

The distortion by the slant of curve should be eliminated in order to measure the maximum deviation angle. This can be done in such manner that the curve of the opposite slant adds to the original slant curve and then, it is averaged. The curve of the opposite slant corresponding to Eq. (22) can be written as

$$\beta''(x,t) = \frac{\delta \cdot (n_2 - n_1)}{2\sqrt{\pi Dt}} \exp\left(-\frac{x^2}{4Dt}\right) - Ax. \qquad (23)$$

Two curves described by Eq. (22) and Eq. (23) are symmetry together with respect to OX axis. Thus, summation of these equations and its average will give the undistorted distribution of deviation angle:

$$\beta(x,t) = \frac{\beta'(x,t) + \beta''(x,t)}{2} = \frac{\delta \cdot (n_2 - n_1)}{2\sqrt{\pi Dt}} \exp\left(-\frac{x^2}{4Dt}\right). \qquad (24)$$

From Eq. (24), the maximum deviation angle at x=0 can be accurately determined. Two images taken at any two times $t_1$ and $t_2$ at diffusion boundary are recorded in a PC and the following figure can be obtained through the process from Eq. (22) to Eq. (24).

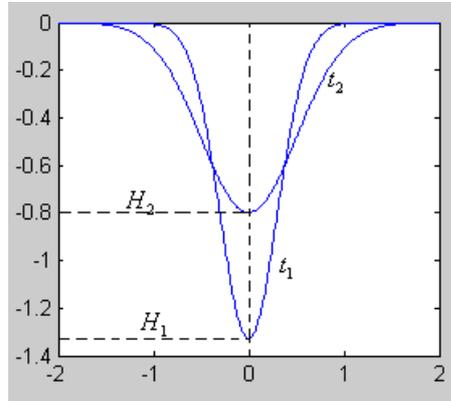

Fig.9.The processed result of two images obtained at any two times at diffusion boundary

From above Fig.9, maximum deviation distances of two curves, $H_1$ and $H_2$ can be found, and finally, diffusion coefficient of solution can be calculated by

$$D = \frac{\delta^2 (n_2 - n_1)^2 L_1^2}{4\pi (t_2 - t_1)} \cdot \left(\frac{1}{H_2^2} - \frac{1}{H_1^2}\right). \qquad (25)$$

Here, parameters on the right hand of Eq. (25) are all the measurable quantities.

## 4. Experimental results and analysis

In the experiment, the diffusion between distilled water and salted solution of 30‰ was measured. The measurements of refractive indices of two solutions were carried out by the use of Abbe refractometer at a temperature of 25℃. Refractive indices of distilled water and salted solution of 30‰ are 1.323 and 1.368, respectively. A dimension of cuvette containing solutions is $40 \times 40 \times 16 mm^3$. Thus, a thickness of solution layer ($\delta$) is 16mm and cuvette-screen separation, $L_1$ is 10cm. The cuvette is first half filled with distilled water and then, the salted solution is carefully injected below by using injector. Radial laser rays are incident at slant angle of 45° on the diffusion boundary, and in this case, the image such as Fig.10 is formed on the screen.

Figs. 10 (a)-(d) are images recorded in **PC** at 5min, 30min, 90min and 140min since the starting of diffusion. As can be seen in Fig. 10, the slant curve is gradually disappeared with time and finally, it is converted as slant line. It means that the concentration difference between two solutions was zero and the diffusion process reached at equilibrium state. In Figs, maximum deviations were determined through Eq. (22) to Eq. (25). The maximum deviation is $H_1 = 17.3mm$ at 5 min since the starting of diffusion and $H_2 = 8.8mm$ at 30min. According to Eq. (25), the averaged value of diffusion coefficient is $D = 2.65 \times 10^{-9} m^2/s$. Compared with the reference datum of $D = 1.554 \times 10^{-9} m^2/s$ [17], a good agreement was not found (due to the different experimental condition, for example, different concentration).

In our case, as the concentration of salted water is higher than that of [17], it is natural that the value of diffusion coefficient is larger. But there is a good coincidence in the tendency. From this, we can conclude that the method we proposed is valid and it gives a much simpler method for the measurement of diffusion coefficient.

## 4. Conclusion

A new simple method for the measurement of diffusion coefficient in transparent solution is proposed. When radial laser rays are incident at constant slope on the diffusion boundary layer, the shape of slant normal distribution function appears. It is difficult to find maximum deviation from this function curve, so **CCD** camera receives this image in real time and a certain transformation is carried out in computer in order to find a new normal distribution function. As a result, maximum deviation can be accurately determined. It is fairly easy to measure diffusivity in such a manner.

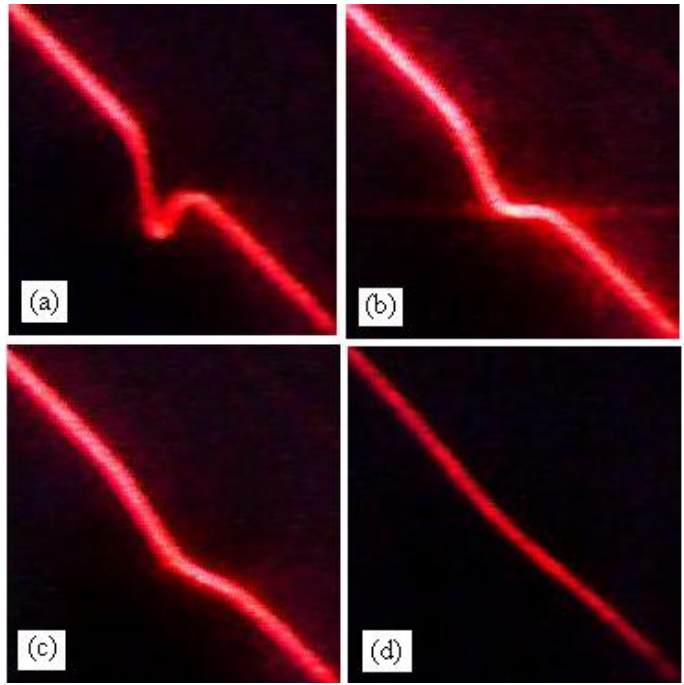

Fig.10.The diffusion process of 30％ salted water into distilled water

    a) after 5min     b) after 30min

    c) after 90min     c) after 140min